# Mechanics and Dynamics of X-Chromosome Pairing at X Inactivation

Antonio Scialdone[1,2], Mario Nicodemi[2,3]*

1 Dipartimento di Scienze Fisiche, Università di Napoli "Federico II", Naples, Italy, 2 INFN, Naples, Italy, 3 Department of Physics and Complexity Science, University of Warwick, Coventry, United Kingdom

## Abstract

At the onset of X-chromosome inactivation, the vital process whereby female mammalian cells equalize X products with respect to males, the X chromosomes are colocalized along their *Xic* (X-inactivation center) regions. The mechanism inducing recognition and pairing of the X's remains, though, elusive. Starting from recent discoveries on the molecular factors and on the DNA sequences (the so-called "pairing sites") involved, we dissect the mechanical basis of *Xic* colocalization by using a statistical physics model. We show that soluble DNA-specific binding molecules, such as those experimentally identified, can be indeed sufficient to induce the spontaneous colocalization of the homologous chromosomes but only when their concentration, or chemical affinity, rises above a threshold value as a consequence of a thermodynamic phase transition. We derive the likelihood of pairing and its probability distribution. Chromosome dynamics has two stages: an initial independent Brownian diffusion followed, after a characteristic time scale, by recognition and pairing. Finally, we investigate the effects of DNA deletion/insertions in the region of pairing sites and compare model predictions to available experimental data.

Funding: No funding was received for this work.

Competing Interests: The authors have declared that no competing interests exist.

* E-mail: mario.nicodemi@na.infn.it

## Introduction

In female mammalian cells one X chromosome has to be inactivated in order to equalize the dosage of X genes products with respect to males [1–3]. Such a phenomenon, known as X-chromosome inactivation (XCI), is regulated by a long region of the X, the *X Inactivation Center* (*Xic*). A crucial initial step that occurs during XCI is the physical colocalization of the two *Xic*'s [4–6]. Disruptions of pairing induce XCI failure and cell death, yet the mechanisms whereby the two *Xic*'s recognize each other and colocalize remain obscure.

In murine embryonic stem cells, pairing occurs within the early days of differentiation. Two major regions of colocalization have been discovered (see Figure 1): a sequence between *Tsix* and *Xite* genes, close to [4,5]; and a segment located several hundred kilobases upstream, named *Xpr* [6], colocalizing independently from *Tsix*/*Xite*. While the specific role of these regions is still under investigation, several details of *Tsix*/*Xite* have been elucidated.

Its colocalization requires some few kilobase long DNA subfragments and a known Zn-finger protein, CTCF, having several DNA binding domains which can bind those subfragments at multiple and clustered sites [4,7] (see Figure 1). Since the inhibition of transcription of *Tsix* and *Xite* disrupts the formation of X–X couples, it has been, thus, proposed that the X chromosome interaction is mediated by a transiently stable "RNA-protein bridge" at these specific *Xic* sites [7].

Importantly, the insertion on autosomes (non sex chromosomes) of the mentioned *Tsix*/*Xite* and *Xpr* segments of *Xic* induces X-autosome pairing [4,6,7]. A still unexplained result is that deletion/insertions, including pairing regions, affect the strength of pairing according to their length, e.g., longer heterozygous deletions exhibit weaker pairing, as in the case of $XX^{\Delta Xite}$ and $XX^{\Delta Tsix}$ entailing the removal of 3.7 and 5.6 kbps respectively [4]. Consistently, longer insertions of *Xic* pairing segments produce stronger X-autosome pairing; and, in females, X-autosome interactions compete with X–X pairing [4]. As deletions of those DNA regions and mutations of CTCF disrupt colocalization, these elements are thought to be necessary components of the pairing machinery. The crucial questions, though, on whether they are sufficient, on the mechanical basis and the physical requirements producing X chromosomes recognition, colocalization and time orchestration, are still unanswered.

To this aim, we investigate a schematic physics model of the molecular elements involved in the pairing of these loci, which includes the general features of DNA sequences (e.g., *Tsix*/*Xite* subfragments) and molecular factors (e.g., CTCF) summarized above. By extensive computer simulations, we show that *Xic* regions can be, indeed, spontaneously colocalized as the result of their interaction with binding soluble molecular factors. Thermodynamics imposes, however, that *Xic*'s do recognize each other and come into physical proximity only if mediator concentration, or their affinity, exceeds a critical value, else they move independently. This grounds on Statistical Mechanics the proposed "RNA-protein" bridge scenario of colocalization [7], by disclosing its physical requirements, it suggests how the cell can regulate it. The model also predicts the kinetics and probability of *Xic* colocalization, which are here compared to available experimental data. Finally, the non trivial effects of deletion/

### Author Summary

Some important cellular processes involve homologous chromosome recognition and pairing. A prominent example is the colocalization of X chromosomes occurring at the onset of X chromosome inactivation, the vital process whereby female mammalian cells silence one of their two X chromosomes to equalize the dosage of X products with respect to males (having just one X). The crucial question on how the Xs recognize each other and come together is, however, still open. Starting from important recent experimental discoveries, we propose a quantitative model, from statistical mechanics, which elucidates the mechanical basis of such phenomena. We demonstrate that a set of soluble molecules binding specific DNA sequences are sufficient to induce recognition and colocalization. This is possible, however, only when their binding energy/concentration exceeds a threshold value, and this suggests how the cell could regulate colocalization. The pairing mechanism that we propose is grounded in general thermodynamic principles, so it could apply to other DNA pairing processes. While we also explore the kinetics of X colocalization, we compare our results to available experimental data and produce testable predictions.

insertions into the X pairing sequences, as much as of chemical modifications of DNA and molecular factors (e.g., CTCF) are investigated.

## Results

### The Model

Since the molecular components of the "RNA-protein bridge" and their DNA binding sites are only partially known (e.g., CTCF and its *Tsix/Xite* binding sites), we consider here a general model, which can accommodate other elements, aiming to depict a broader scenario. In our physics model, the two *Xic* segments involved in the process are represented as two directed polymers of $n$ beads (see Figure 2 and Methods), a well established model of polymer physics [8], while, for sake of simplicity, the rest of the X chromosomes is neglected. Along the polymers, a subset of beads acts as binding sites (BSs, green beads in Figure 2) for molecular factors (MFs). They are contiguous in resemblance of the clustered DNA sites of CTCF [4,7,9], and their number, $n_0$, is chosen to be $n_0 = 24$, i.e., of the order of magnitude of CTCF known sites in the *Tsix/Xite* region [9], although, it will be varied to illustrate the effects of deletions (see Figures 1 and 2, Methods and below).

The MFs represent specific complexes of soluble binders for *Tsix/Xite* or *Xpr*, such as RNAs or CTCF; and we name $c$ their concentration. Since CTCF has many DNA binding domains, we suppose that each MF can bind both polymers at the same time. Analogously, a MF could be a complex including two (or more) molecules able to bind one DNA site each. Different binding sites are likely to have different chemical affinities in reality, nevertheless, we suppose they are all equal to an average value, $E$, and we mostly focus on the case where $E$ is in the range of weak biochemical energies, as expected from the CTCF example. MFs and the two DNA segments float in a box of given size (see Figure 2 and Methods). By extensive Monte Carlo computer simulations [10], the thermodynamics and, later, the dynamics of the system are investigated.

### Chromosome Spontaneous Colocalization

The 'pairing state' of the *Xic* segments is monitored by measuring the average fraction, $p$, of colocalized *Xic*'s, i.e., the fraction of couples whose equilibrium mean distance is less than 10% of the linear size of the including volume.

In presence of a given concentration, $c$, of MFs a bridge between the two polymers could be formed by chance when a MF binds both simultaneously. As a single bridging event is statistically unlikely and short lived, the degree of pairing between the *Xic*'s is expected to be stronger the higher $c$. However, a threshold behaviour exists. Figure 3 shows the thermodynamic equilibrium value of $p$ as function of $c$ (for $E = 1.2\ kT$): below a threshold, $c^* \simeq 2.3\%$ (defined by the inflection point of $p(c)$), $p$ is practically zero irrespective of $c$. Such a region is the 'Brownian phase' where molecules are unable to form thermodynamically stable bridges and chromosomes move independently (a typical configuration is depicted in Figure 2A). Conversely, if $c$ is above threshold, $p$ rapidly approaches 100%, signaling the transition to a different regime, the 'pairing phase': here chromosomes have spontaneously, and ineluctably, recognized and paired by the effects of a stable effective attraction induced by MFs (a configuration is shown in Figure 2C). The colocalization process is, though, fully reversible by reduction of MF concentration below threshold. Around $c^*$, a narrow crossover region exists between these two phases where $p(c)$ is significantly different from zero, but still well below 100%, as couples of chromosomes are continuously formed and disrupted (see the configuration in Figure 2B). In the thermodynamic limit (i.e., for an infinitely large system), the inflection point, $c^*$, corresponds to a phase transition. For a finite sized system, as the one simulated here, $p(c)$ is well fitted by an exponential:

$$p(c) = 1 - \exp\left[-(c/c_0)^\beta\right]$$

where $\beta$ is a fitting parameter (for the case discussed in Figure 3, we find $\beta = 4.7$) and $c_0$ is proportional to the threshold value, $c^*$, found above $\left(c^* = c_0 \left(\frac{\beta-1}{\beta}\right)^{1/\beta}\right)$.

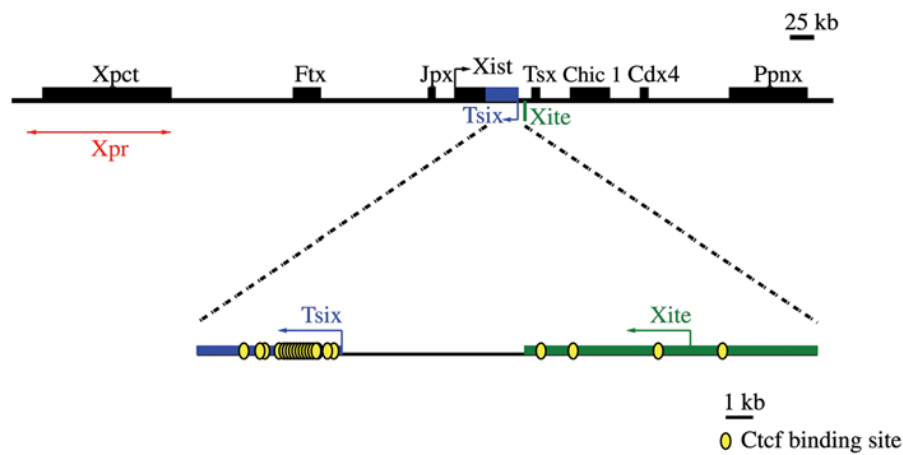

**Figure 1. Diagram of the *Xic* region involved in X chromosome pairing.** The location of *Xpr* [6] and *Tsix/Xite* [4,5], the regions involved in pairing at the onset of X-Chromosome Inactivation (XCI), is mapped within the X-Inactivation center (*Xic*). The red line with arrows highlights the area where *Xpr* has been localized [6]. The enlargement of the *Tsix/Xite* region reports the discovered binding sites for CTCF [7].
doi:10.1371/journal.pcbi.1000244.g001

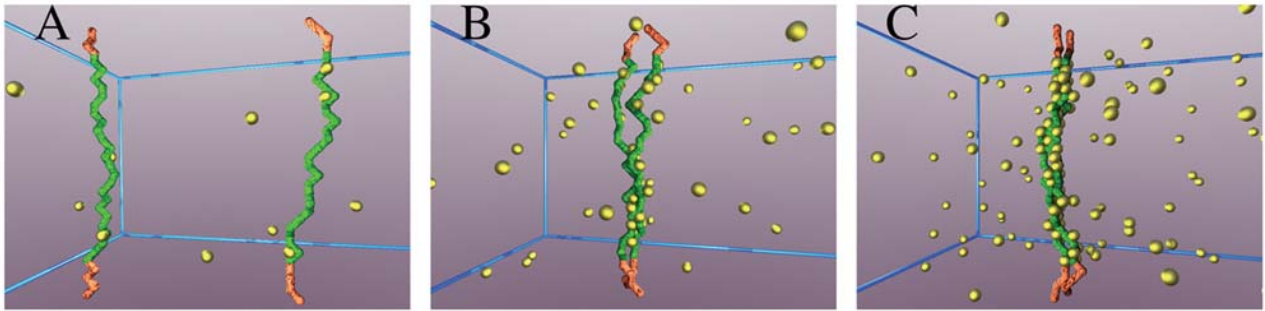

**Figure 2. Typical equilibrium configurations.** Pictures of typical configurations of our model system at thermodynamic equilibrium (here $E = 1.2 \, kT$). (A) Polymers conformation for a value of the concentration of molecular factors (MFs) $c = 0.3\%$ (Brownian phase, see Figure 4), (B) for $c = 2.5\%$ (crossover region), (C) for $c = 5\%$ (Pairing phase). The polymers, representing *Xic* segments responsible for pairing, are formed by a set of linked beads (not visible because of magnification); green beads are the binding sites (BSs) interacting with the floating molecular factors (MFs, yellow beads). The BSs form a cluster of $n_0 = 24$ sites, which is of the order of magnitude of the clustered CTCF binding sites found in the *Tsix/Xite* region (see Figure 1). MFs can bind more than a single BS at the same time, as much as CTCF molecules which have multiple DNA binding domains.
doi:10.1371/journal.pcbi.1000244.g002

For a given value of $c$, the route to colocalization could be taken, analogously, by increasing $E$, as a result, for instance, of modifications of the DNA binding regions or of mediating molecular complexes. Under this different path, a threshold exists as well, as summarized in the phase diagram of Figure 4, showing the equilibrium pairing state of the *Xic*'s in the concentration-energy plane, $(c, E)$, along with the transition line, $c^*(E)$, between the two phases. The existence of the threshold line, $c^*(E)$, has its roots in a thermodynamic phase transition occurring in the system, where the energy gain resulting from pairing compensates the corresponding entropy loss [11]. In particular, we find a power law behavior for the function $c^*(E)$ at small $E$ (superimposed fit):

$$c^*(E) \sim (E - E_{min})^{-\nu}$$

where the exponent is $\nu \sim 4$, and $E_{min} \simeq 0.7 \, kT$ is a minimal threshold energy below which no pairing transition is possible. At higher $E$ an exponential fit of $c^*(E)$ works as well. The phase diagram of Figure 4 gives precise constraints to the admissible values of $c$ and $E$ to attain colocalization. On the other hand, colocalization is found in a broad range of weak biochemical binding energies and concentrations, pointing out that the recognition/pairing mechanism here envisaged is robust, irrespective of the ultimate biochemical details of the "RNA-protein bridge" and its binding sites.

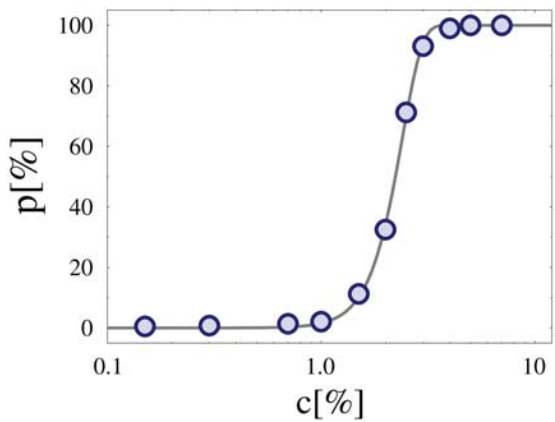

**Figure 3. Equilibrium state as function of $c$.** The equilibrium value of the fraction of paired chromosomes, $p$, is plotted as function of the concentration, $c$, of binding molecular factors, for a given value of their affinity, $E$ (here, $E = 1.2 \, kT$). When the concentration is below a threshold value $c^* \simeq 2.3\%$, no stable pairing is observed ($p \sim 0$) and the chromosomes randomly float away from each other ('Brownian phase'). Above threshold, $p$ saturates to 100%, as a phase transition occurs (to the 'Pairing phase') and chromosomes spontaneously colocalize, their driving force being an effective attraction of thermodynamics origin.
doi:10.1371/journal.pcbi.1000244.g003

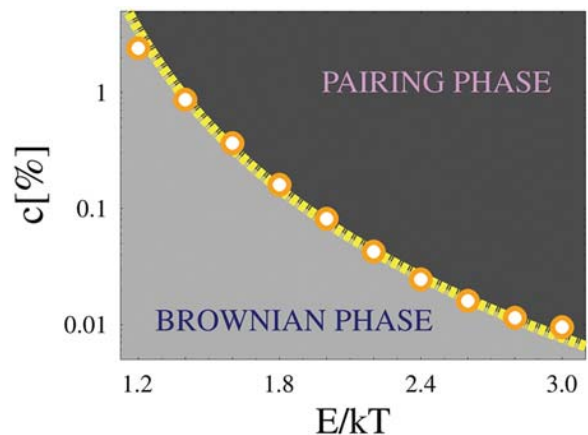

**Figure 4. Phase diagram.** The diagram shows the thermodynamic equilibrium state of the system in the $(E, c)$ plane, for a range of typical biochemical binding energies, $E$, and concentrations, $c$. Circles mark the line $c^*(E)$ delimiting the transition from the Brownian phase, where chromosomes diffuse independently, to the Pairing phase, where chromosomes are juxtaposed (the superimposed fit is a power law).
doi:10.1371/journal.pcbi.1000244.g004

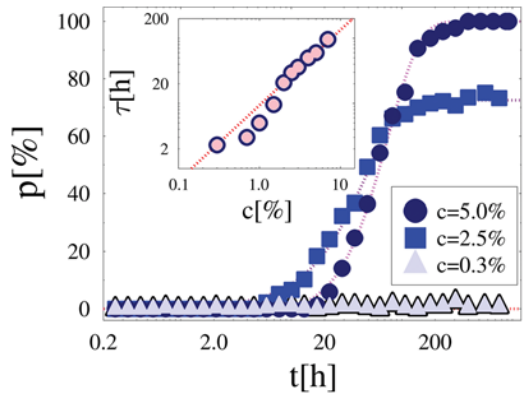

**Figure 5. System Dynamics.** The average fraction of paired chromosomes, $p$, is plotted as a function of time, $t$, for three values of the concentration of molecular factors, $c$ (here $E = 1.2$ $kT$), belonging to three different regimes: Brownian $c = 0.3\%$; crossover $c = 2.5\%$; Pairing $c = 5\%$. After an initial diffusive behavior, chromosomes attain their equilibrium pairing state exponentially in time (superimposed fit: $p(t) \propto [1 - \exp(-t/\tau)]$). Inset: The average time scale, $\tau$, to attain the equilibrium pairing state is plotted as function of $c$ (for $E = 1.2$ $kT$). $\tau$ increases with $c$ because the higher $c$, the higher is the average number of molecules bound to DNA and, consequently, proportionally lower the $Xic$ diffusion constant. The superimposed fit is a linear function.
doi:10.1371/journal.pcbi.1000244.g005

### The Dynamics of Colocalization

To analyze the system dynamics, in our simulations MFs are initially randomly distributed in the enclosing volume and the two DNA segments start from the maximal possible distance from each other, so, at time $t = 0$ the pairing fraction is $p = 0$. Figure 5 reports $p(t)$, i.e., the change in time of $p$, for three values of the concentration which belong to the different regimes discussed above (here, $E = 1.2$ $kT$). For $c = 0.3\%$, i.e., below threshold, $p(t)$ never rises above the background zero value. By increasing $c$ to $c = 2.5\%$ and $5\%$, close to and above the threshold, chromosomes "sense" each other [6], as $p(t)$ grows to its non-zero equilibrium value, $p(c)$, seen before. After an initial Brownian regime, where $p(t) \propto t$, at long times an exponential dynamics is recorded (superimposed fit in Figure 5):

$$p(t) \simeq p(c)\left(1 - e^{-t/\tau}\right)$$

where $p(c)$ is the equilibrium value discussed above. The parameter $\tau$, a function of $c$ and $E$, is a measure of the average time to attain equilibrium (and pairing for $c > c^*$) [Note: a full fit function for $p(t)$ is: $p(t) = p(c) - \left[p(c) + \frac{\gamma t}{1+\delta \cdot t}\right]e^{-t/\tau}$, where $\gamma$ and $\delta$ (such as $\tau$) are fit parameters depending on the values of $c$ and $E$. This fits our data very well, however, we prefer to report the simpler one-parameter exponential fit which is sufficient to estimate the characteristic time scale, $\tau$. The two fits provide approximately equal values of $\tau$].

Interestingly, $\tau$ increases when the number of MFs increases, approximately linearly in $c$ (see inset Figure 5 and superimposed fit):

$$\tau(c) = \tau(0)c^\alpha$$

where the fit parameters are $\tau(0) = 15$ $h$ and $\alpha = 1.1$. The rationale for such a behavior is that when $Xic$ segments are bound by MFs their effective diffusion constant is proportionally reduced and so, while they will be eventually colocalized, the time to equilibrium is longer the higher the concentration of MFs.

In our model we consider only short segments of the $Xic$'s. For that reason, a comparatively strong effect of MF concentration, $c$, on $\tau$ is found. In real systems, the situation is more complicated, also because other phenomena may affect the kinetics of chromosomes. Nevertheless, a dependence of the diffusion constant on the molecular factor concentration should be observed locally, at the scale of the pairing sequences.

### The Distribution of Chromosome Distance

Further insight and direct comparisons to known experimental results [4–6] are obtained by studying the dynamics of the probability distribution, $\mathcal{F}$, of the normalized distance, $ND$, between the X chromosomes ($ND$ is normalized by the system volume linear size, so $0 < ND \leq 1$). In these simulations, the two DNA segments are initially randomly scattered across the lattice, as much as the MFs. The typical shape of the function $\mathcal{F}(ND)$, in the region where pairing occurs, is reported in Figure 6A, at two time frames. At $t = 0$, $\mathcal{F}$ has the same shape, and approximately the same mean value, of the normalized $Xic$ distance distribution measured in mouse embryonic stem cells at the beginning of differentiation ([4] and inset of Figure 6A). In mice, $Tsix/Xite$ pairing is observed approximately within day two from differentiation [4–7]. In our simulations, a peak in $\mathcal{F}(ND < 0.1)$ grows in time and after around 48 h saturates to its final value (for $c = 5\%$ and $E = 1$ $kT$). The final distribution is bimodal: superimposed to the peak in $\mathcal{F}(ND < 0.1)$, which corresponds to the chromosomes bridged by MFs, there is a much broader distribution (very similar to the initial random one, with a peak approximately centered in $ND \simeq 0.5$) which corresponds to independently floating chromosomes. The cumulative frequency distribution of chromosomes with $ND$ below 0.1, i.e., 'paired chromosomes', plotted in Figure 6B, shows the growth of the component of bridged chromosomes in the statistical population. The overall shape of the distribution, $\mathcal{F}(ND)$, and its change with time, found in our simulations reproduce qualitatively very well those observed in the experiments (see inset of Figure 6A). This is intriguing when considering the simplicity of our model and the fact that we use just simple reasonable guess values for binding energy, $E$, and kinetic constant, $r_0$ (which have still to be experimentally found). A number of further complexities are present in real experiments (e.g., averages over non uniform populations of cells, difficulties in resolving neighboring fluorescent spots, etc.), which could explain discrepancies in the details.

To estimate the fraction of pairing events in a given time window, it is also important to calculate the distribution of 'collision' times: we measured the time, $t_{collision}$, needed by a chromosome to encounter for the first time the other, i.e., to be located within a normalized distance, $ND$, less than 0.1 from it; Figure 6C illustrates the probability distribution, $\mathcal{P}$, of $t_{collision}$ (for the same parameter values of Figure 6A and 6B). Note that $t_{collision}$ is the time for just a 'collision' to occur, which could either result in a stable pairing or not. $\mathcal{P}$ is approximately exponential in $t_{collision}$:

$$\mathcal{P}(t_{collision}) = \mathcal{P}_0 \exp(-t_{collision}/t_0)$$

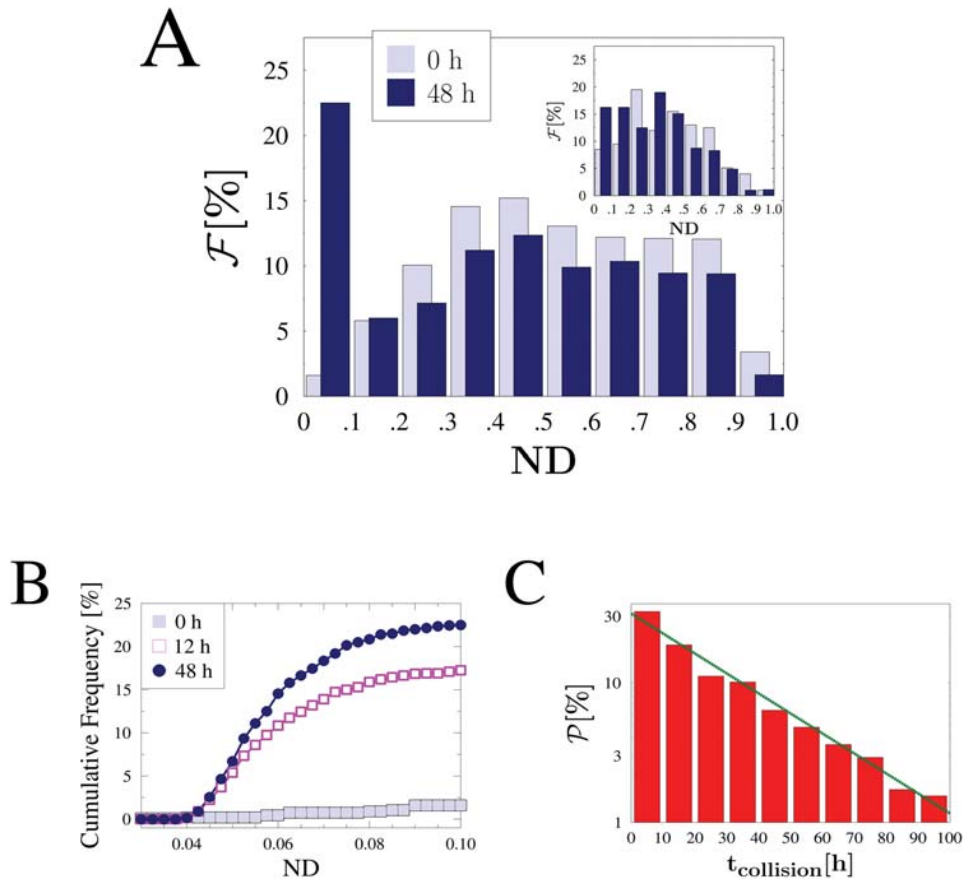

**Figure 6. Distance and collision times distribution.** (A) The distribution $\mathcal{F}$ of the normalized distance, $ND$ ($0<ND\leq 1$), between the two X chromosomes is plotted at two time frames (in the phase where pairing occurs, here $c=5\%$, $E=1$ $kT$). The initial distribution corresponds to randomly located chromosome positions ($t=0$ h); while colocalization progresses a peak in $\mathcal{F}(ND<0.1)$ becomes visible and saturates at 48 h. In the inset the corresponding experimental data (from [4]) are reported. (B) The cumulative frequency distribution of 'paired chromosomes'(i.e., having $ND<0.1$), under the same conditions of (A), is shown. (C) Probability distribution $\mathcal{P}$ of the time $t_{collision}$ required by a chromosome to encounter for the first time the other (i.e., to be located within a normalized distance, $ND$, less than 0.1 from it) with the same values for $E$ and $c$ used in (A) and (B). An exponential behaviour is found (superimposed fit).
doi:10.1371/journal.pcbi.1000244.g006

$\mathcal{P}_0$ and $t_0$ being fit parameters (for the case in Figure 6C we find $\mathcal{P}_0=31.5\%$ and $t_0=30$ h).

### Heterozygous Deletions

Finally, we also explored the effects on pairing of heterozygous deletions, an issue of practical relevance to experimental studies. We consider the case where the BSs on one $Xic$ are reduced to a fraction, $f$, of their original number $n_0$ (for the same MF concentration, $c=5\%$, and affinity $E=1.2$ $kT$, mostly discussed before). While a reduction of the pairing fraction, $p$, is expected in presence of a deletion, we find that the equilibrium value of $p$ has a non-linear behavior in $f$, a sigmoid with a threshold at $f^*\sim 50\%$ (see Figure 7): short deletions (say, preserving a fraction of BSs $f\gtrsim 70\%$) do not result in a relevant reduction of $p$, while pairing is completely lost as soon as $f$ gets smaller than about 30%. The sigmoid behavior stems from the non trivial thermodynamic origin of the MF mediated effective attraction between $Xic$'s. The threshold value, $f^*$, is a decreasing function of $E$ and $c$. While similar considerations to those mentioned when discussing $p(c)$ apply, we find that an exponential fits $p(f)$ (superimposed line in Figure 7): $p(f)=1-\exp[-(f/f_p)^\lambda]$, where $\lambda$ is a fitting parameter (for the case discussed in Figure 7, we find $\lambda=4.2$) and $f_p$ is proportional to the threshold $f^*$ discussed above $\left(f^*=f_p\left(\frac{\lambda-1}{\lambda}\right)^{1/\lambda}\right)$. While these results rationalize the observed length dependent effects of $Tsix/Xite$ deletions [4], the predicted behavior of $p(f)$ could be experimentally tested.

Interestingly, the time to approach the equilibrium state, $\tau$, is smaller the longer the deletion, as shown in the inset of Figure 7. Such a seemingly counterintuitive result stems from the fact that the smaller $f$, the smaller is the number of MFs attached to Xic segments and, thus, larger their effective diffusion constant (see above). The function $\tau(f)$ is well fitted by: $\tau(f)=\tau(0)+\Delta\tau[1-e^{-(f/f_\tau)^\eta}]$; for the case shown in the inset of Figure 7, the fit parameters are $\tau(0)=3.6$ h, $\Delta\tau=56.4$ h, $f_\tau=59\%$ and $\eta=3.5$, but they are functions of $E$ and $c$ (as discussed above). The increasing behaviour of $\tau$ as function of $f$ results in a non-trivial prediction: the removal of a fraction of BSs within a chromosome should speed $Xic$ pairing with respect to the Wild Type case, although, the overall fraction of paired chromosomes would be reduced. As seen in the inset of Figure 5, an analogous phenomenon would be observed by decreasing the concentration of MFs.

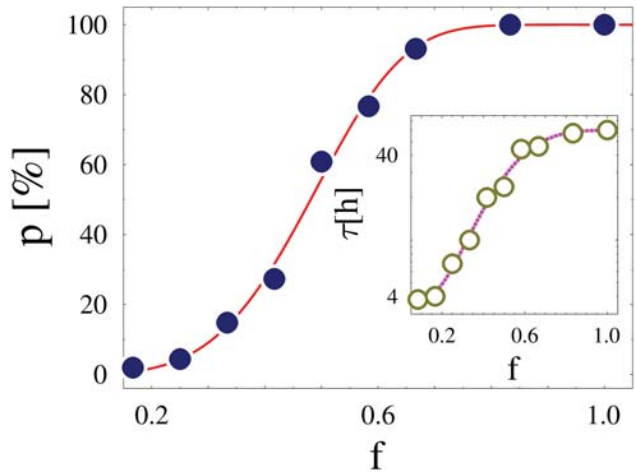

**Figure 7. Binding sites deletions.** The figure shows the pairing fraction, *p*, in heterozygous deletions, as a function of the remaining fraction, *f*, of original binding sites. In the 'Wild Type' case ($f=1$) the system is chosen to be in the 'Pairing phase' (here $c=5\%$, $E=1.2\ kT$) and the equilibrium value of the fraction of paired chromosomes is $p=100\%$. The pairing fraction, *p*, has a non linear behavior as function of *f*, with a crossover region around $f\sim50\%$. Short deletions, preserving a large fraction of BSs, say, $f>70\%$, have tiny effects on the pairing fraction, while deletions with $f<30\%$ erase pairing. Inset: The average time, $\tau$, to approach the equilibrium pairing state is plotted as function of *f*. When *f* is reduced, $\tau$ is shorter, since less MFs are bound to Xic's which, in turn, have an higher effective diffusion constant.
doi:10.1371/journal.pcbi.1000244.g007

## Discussion

Our polymer physics model gives quantitative foundations to the 'molecular bridge' scenario [7] for X chromosome colocalization at XCI in the *Tsix/Xite* locus. In a process involving a phase transition from a 'Brownian' to a 'Pairing phase', it shows that clusters of DNA sites can recognize each other on different chromosomes and come in physical proximity by interacting with diffusible binding molecular factors. Thermodynamics imposes, however, that X colocalization can be spontaneously attained only when the concentration, or affinity, of molecules is above a critical value. Weak biochemical interactions can be sufficient for pairing (when *c* is above threshold), as much as higher energies, e.g., related to specificity of binding. In real cells, a pairing initially based on weak interactions would have the advantage of avoiding topological entanglement by leaving space to adjustments.

From our calculated values of threshold concentrations we can roughly estimate the correponding molecular concentrations in real nuclei: in our model *c* is the number of molecules per lattice site, so the number of molecules per unit volume is $c/d_0^3$, where $d_0$ is the linear lattice spacing constant. If this quantity is divided by the Avogadro number $\mathcal{N}_A$, the molar concentration $\rho$ is obtained. Results from Figure 4 suggest that a typical value of threshold concentration could be around $c=0.1\%$. Under the rough assumption that $d_0$ is a couple of orders of magnitude smaller than the nucleus diameter (i.e., $d_0\sim10\ nm$), a typical threshold molar concentration would be $\rho\sim1\ \mu mole/litre$. In the case of CTCF molecule, this corresponds to a mass concentration of $\sim0.1\ mg/ml$, a value which is compatible with typical values of nuclear protein experimental concentrations (see [12,13]). The above calculation is very rough (e.g., the critical value for concentration strongly depends on the value of binding energy *E*, see Figure 4, and we made just a reasonable assumption about the value of $d_0$), but may help to further bridge our study with biological investigations.

Our picture explains, thus, how well described cell strategies to change genomic architecture (upregulation of a DNA binding protein or modification of chromatin structure) can act to produce *Xic* colocalization. Although many features of X pairing at XCI are still unknown, the model describes the overall experimental scenario, including the observed effects of deletions/insertions length on the strength of pairing [4] and provides, for the first time, a precise description of X colocalization kinetics. A quantitative picture emerges of the physical mechanisms underlying the early stages of XCI [14,15], along with non-trivial predictions on colocalization kinetics and on the outcome of genetic deletions/insertions within the *Tsix/Xite* region, such as the presence of threshold effects or the changes in the pairing kinetics. The present version of the 'molecular bridge model' could, thus, guide the design of future experiments to elucidate X pairing at XCI. Open questions regard the specific molecular differences between the two pairing regions *Tsix/Xite* and *Xpr*, and, more generally, whether the thermodynamically robust mechanisms described here apply to other cell processes involving recognition and pairing of DNA sequences [11,16–22].

## Methods

We describe DNA segments via a standard model of polymer physics [8] as directed chains of $n=32$ beads (see Figure 2); $n_0=24$ of them are binding sites for molecular factors (MFs). For computational purposes, in our Monte Carlo computer simulations [10] polymers and MFs are constrained to move on the vertexes of a square lattice of linear sizes $L_x=2L$, $L_y=L$ and $L_z=L$ (in units of $d_0$, the length of a single BS) with $L=32$, under periodic boundary conditions. These choices of parameters results from comparisons to experimental data on CTCF and its DNA binding sites at the *Tsix/Xite* locus [4,7] (e.g., we choose the BS number close to currently known number of CTCF binding sites in the *Tsix/Xite* regions xci_donohoe) and from computational feasibility requirements. While the robustness of our model is well established in polymer physics [8], we checked that our general results are unchanged by using different values of these parameters (we tested lattice sizes up to $L=128$, and combinations of $n$ and $n_0$ as large as 128). Real DNA pairing loci of the *Xic* are likely to differ in size (i.e., $n_0$) and arrangement of their binding sites with respect to the simple case dealt with here. In the light of our investigation, such differences can affect the details of their behaviors, but the general picture we depict is not altered. In cases where detailed data on binding sequences and regulator chemistry is available, such information could be easily taken into account in the model to produce very detailed quantitative descriptions.

Polymers beads diffuse without overlapping and under the constraint that the 'chromosome' is not broken, i.e., that two proximal beads must be on next or nearest next neighboring sites on the lattice. MFs diffuse as well without overlapping. A bond can be formed between a MF and a BS only if they are on neighboring sites. Sites and particles move to a neighboring lattice vertex with a probability proportional to the Arrhenius factor $r_0 e^{-\Delta E/kT}$, where $\Delta E$ the energy barrier in the move and $r_0$ is the reaction kinetic rate. The conversion factor from Monte Carlo time unit to real time is established [10] by imposing that chemical reactions involved in bond formation have the same rate of occurrence in Monte Carlo dynamics and in real dynamics. Since the exact value of $r_0$ is unknown in the present case, we use $r_0=15\ s^{-1}$, a typical value for biochemical reactions. Changes to $r_0$ would rescale, inversely, the time axes in our Figures 5, 6, and 7. Our results remain essentially unchanged within a broad range of biochemical values of $E$ (see Figure 4 and [11]).

In our description, we use the approximation whereby the X segments responsible for pairing are represented as directed polymers. The advantage of such an approximation is to permit comparatively faster simulations for our many body system (including a large number of degrees of freedom). It doesn't affect, however, the overall system behaviour. In fact, in our model, pairing is based on a robust thermodynamic mechanism: when the concentration of MFs (or their chemical affinity) increases above a threshold value, the energy gain resulting from bond formations between paired chromosome sites compensates the corresponding entropic loss due to pairing. Thus, in absence of directed polymer constraint, chromosomes will pair as well, as a consequence of such a free energy minimization mechanism. If the constraint is released, however, polymer sequences would pair in more disordered configurations, not perfectly aligned as in the case considered here.

The distance between the polymers in a given configuration is evaluated by averaging the distances between beads, at the same 'height' $z$, belonging to different polymers.

Averages are over up to 2000 runs from different initial configurations.

## Author Contributions